# Beyond Metadata: The Role of DOI, ORCID, and ROR in Shaping Transparent and Interoperable Library Systems

**Snehasish Paul**
*Ph.D. Scholar, Department of Library and Information Science, University of Delhi, Delhi- 110007, India*
*E-mail: snehasishpaulas98@gmail.com*

**Abstract**

The digital transformation of scholarly communication has changed the way libraries manage, preserve scholarly research and share it with the public. This research looks at three persistent identifier (PID) systems: the Digital Object Identifier (DOI), the Open Researcher and Contributor ID (ORCID), and the Research Organization Registry (ROR), which are necessary for a transparent and interoperable library infrastructure. Evidence from 2019–2026 shows how PIDs have changed from metadata tags to machine-actionable connective tissue that links researchers, institutions, publications, datasets, and funding. Findings from implementation studies at the global level show very diverse adoption of ORCIDs, between 41% and 89% among German research organizations, yet continued issues with metadata quality and PID literacy. This research identifies promising practices in Europe and Latin America, examines barriers, and proposes measures for libraries, universities, policymakers, and Library and Information science professionals.

## 1. Introduction

### 1.1 From Tradition Metadata to Smart Linked Metadata

Library and information systems have evolved from isolated repositories of institutional metadata to globally networked systems. The previous cataloging styles were either not consistent in naming, machine-readable, or used proprietary identification (Meadows et al., 2019). In essence, persistent identifier systems represent a move towards "smart linked metadata" that is structured, machine-actionable and enables automated discovery, verification and integration with heterogeneous systems. PID-enabled records are different from conventional metadata because they create a web of relationships among research objects, their creators and the producing organizations in a semantic manner.

### 1.2 The Imperative for Transparency and Interoperability

Library systems must be open, verifiable, and interoperable in today's research environments. Funders want verifiable attribution; institutions need accurate data about research performance; researchers want global discoverability. With the rise of all kinds of depositaries and open access, there is an increasing need for identifiers that work across systems and jurisdictions (Haak et al., 2012). When identifiers aren't resolved and shared, metadata, even if/where well curated, remains an island; viewable in one system, but not or ambiguously so in many others.

Three PIDs have emerged as the core infrastructure responding to these needs: the DOI for digital objects, ORCID for researchers, and ROR for organizations. The PID ecosystem is defined in this research as the combined structure upon which all modern library systems and open science infrastructure depend. This research outlines the ecosystem's conceptual foundations, practical implementations, documented benefits, persistent hurdles and future trajectory, drawing on existing empirical studies and global case examples from 2019 to 2026.

## 2. Conceptual Background

### 2.1 Defining Persistent Identifiers

A persistent identifier is a reference that does not change when the location, owner, and/or format of a digital resource changes. When content moves, URLs may break. PIDs use resolver systems that link users to the resource's current location. Preservation into the long term of the integrity of citations, attribution and institutional records, etc., is shown in scholarly communication (Fenner et al., 2019).

### 2.2 DOI, ORCID, dan ROR: Core Functions and Recent Scale

The international DOI Foundation manages the DOI. It is the Digital Object Identifier, which can be registered by agencies like CrossRef and DataCite. Most importantly, it gives a unique identification to scholarly publications, datasets, and other research. As of 2024, its 290 million DOIs on record make it the world's most used scholarly identifier (Crossref et al., 2023).

ORCID (Open Researcher and Contributor ID) lends a 16-digit identifying number to an individual researcher to disambiguate them from their name variants during a career or institutional changes. In 2024, over 21 million iDs were issued by ORCID, with more than 1,400 member organizations, including publishers, funders and universities, currently active using ORCID (ORCID, 2024). The Research Organization Registry (ROR) provides identifiers for research organizations. First launched in 2019, ROR backers expect it to standardize affiliation metadata (ROR, 2024). By 2024, ROR has registered over 105000 organizations.

### 2.3 The Ecosystem of Persistent Identifiers (PID)

**Table 1:** Comparative Overview of DOI, ORCID, and ROR

| Dimension | DOI | ORCID | ROR |
|---|---|---|---|
| **What it identifies** | Digital objects (articles, datasets, books, software) | Individual researchers and contributors | Research organizations and institutions |
| **Governed by** | International DOI Foundation / Crossref / DataCite | ORCID Inc. (non-profit) | Community governed (California Digital Library, Crossref, DataCite) |
| **Launched** | 2000 | 2012 | 2019 |
| **Format** | 10.XXXX/XXXXX | 0000-0000-0000-0000 | https://ror.org/XXXXXXX |
| **Scale (2024–2025)** | 290+ million registered; 92M DataCite DOIs in OpenAlex | 21+ million iDs issued; 58.8% Czech CRIS coverage; 190K Spanish profiles | 105,000+ organizations; institutional hardcoding at UBC |
| **Primary users** | Publishers, libraries, repositories, data centers | Researchers, publishers, funders, and universities | Funders, publishers, CRIS systems, repositories |
| **Core benefit** | Persistent linking and citation | Author disambiguation and | Standardized affiliation and institutional analytics |

|  | integrity; dataset discoverability | attribution; profile aggregation |  |
|---|---|---|---|
| **Key challenge** | Metadata completeness (54.6% DOI coverage in OA diamond journals); link rot prevention | Low adoption in developing countries, profile maintenance, and uneven national coverage | Awareness; integration with legacy HR/CRIS systems; multilingual disambiguation |
| **Recent developments (2023–2025)** | Nelson Memo re-curation mandates; OpenAlex integration; repository workflow automation | National monitoring (Czech, Spain); CRIS integration; funder mandates | UBC institutional implementation; OpenAlex ML training; AffilGood disambiguation tools |

The three identifiers are complementary: DOI identifies what was produced, ORCID identifies who produced it, and ROR identifies where it was produced. By connecting research outputs with their creators and institutions, a machine-readable scholarly graph is created for the first time. The interface will enable automated reporting, discovery and research assessment at scale.

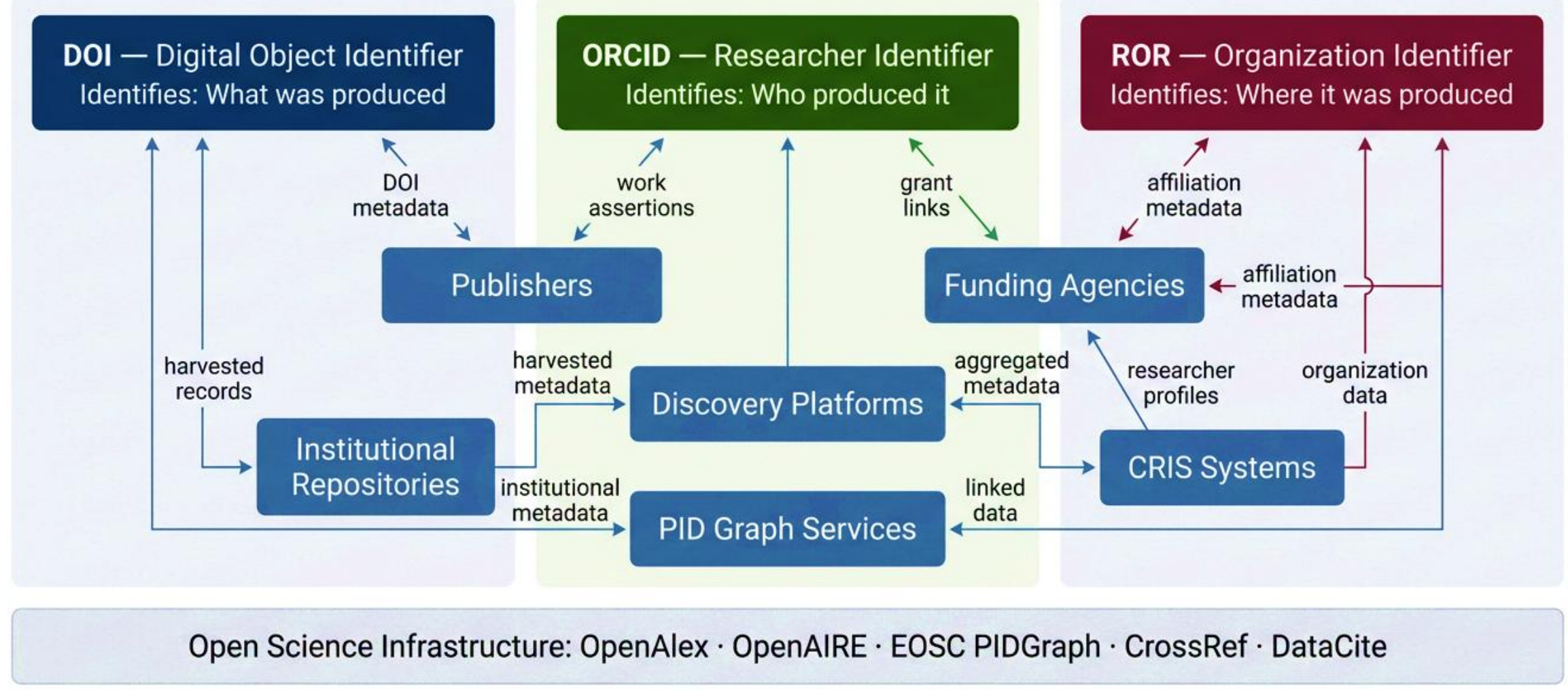

**Figure 1:** The PID Ecosystem: Integrating DOI, ORCID, and ROR
**Source(s):** Created by the author

## 3. DOI in Library and Research Systems

### 3.1 Citation Linking and Long-Term Access

The core function of DOI in library systems is linking citations efficiently. When a reference list contains a DOI, it resolves to the current location of the work, regardless of publisher or platform moves. CrossRef's cited-by linking service enables the interconnection of citing and cited works of publishers, which has resulted in a citation graph. Libraries and discovery systems can programmatically query the citation graph (Hendricks et al., 2020). Tools like Web of Science and Google Scholar, along with Scopus, which uses an internal algorithm, rely on DOI-based matching to build citation metrics.

DOIs that are used for research data registration that were reserved through DataCite. The metadata scheme of DataCite mandates structural information about data creators, related publications and funding sources. Libraries can build data catalogues rich in context using such metadata thoroughly identified and structured. (Kiefer, 2015) asserted that the DOI integrating with preservation services, CLOCKSS, and Portico means that even if the publisher stops operating, the content with a DOI will be maintained through archived copies.

### 3.2 Digital Preservation and Research Visibility

DOI registration creates a baseline preservation record, independent of platform, through a metadata deposit. In making the Libraries a DataCite member, libraries can mint DOIs directly for the institution's own outputs like theses, reports, data sets and grey literature, which gives persistent identification outside the commercially published material. The ability to achieve this development completes the signifier that can be globally resolvable, which is important for institutional repositories. Because locally produced scholarship is under-identified globally.

Enhancing research visibility can happen through DOI-based indexing. The publications that have a DOI are more easily harvested by aggregators, more accurately matched in citation analyses, and more likely to appear in discovery systems. It has been noted that articles with a DOI receive higher citations due to better discoverability and more accurate referencing (Brase et al., 2015). As repositories and institutional publishing programs, libraries have strategic roles in the DOI registration ecosystem as active DOI registration agents and not merely as DOI consumers.

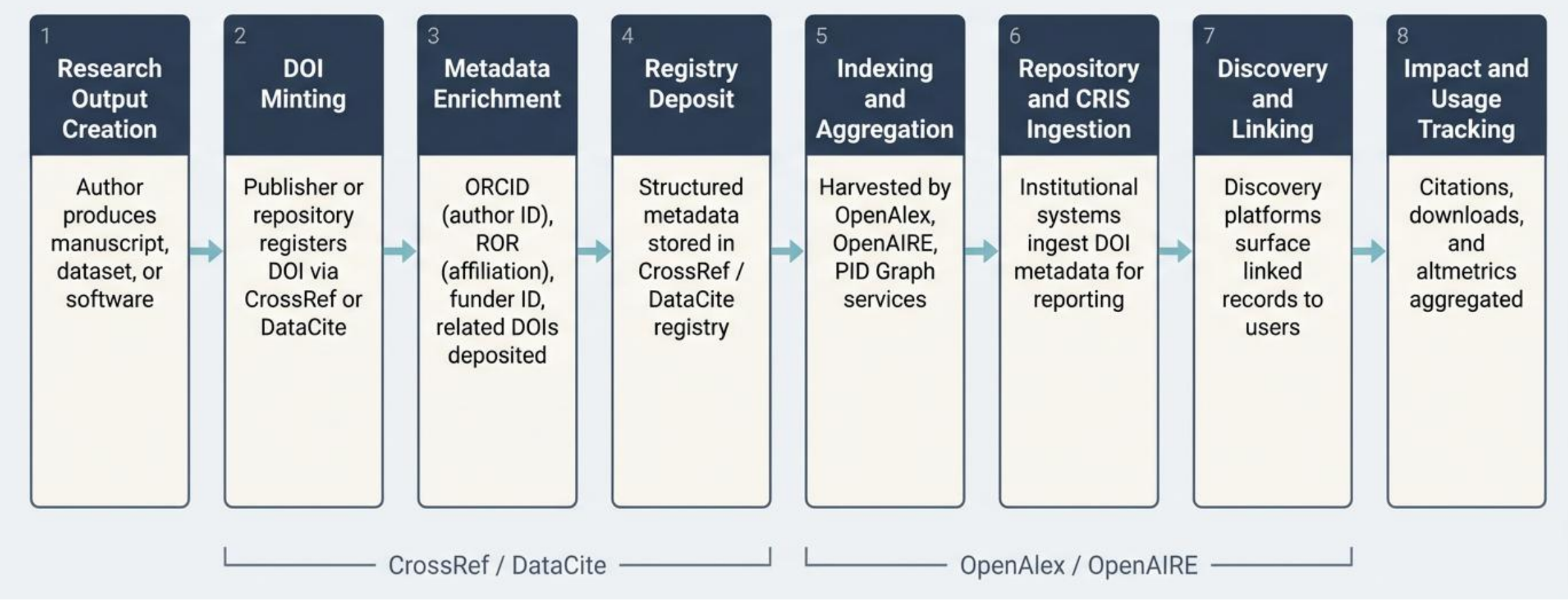

**Figure 2:** The DOI Lifecycle in Scholarly Communication
**Source(s):** Created by the author

## 4. ORCID for Researcher Identity Management

### 4.1 Author Disambiguation and Attribution

The persistent problem of author name ambiguity is a threat in scholarly communication. Attributing works to their rightful creators without a doubt must be done with care as many common names, transliteration differences, name changes after marriage or migration, and a global increase in research output. As Haak et al., (2012) stated, ORCID addresses this issue by giving researchers a persistent identifier that stays with them throughout their careers, regardless of institution, name or publication.

An ORCID record is a researcher's authoritative scholarly profile that brings together a person's publications, datasets, grants, peer review and employment in one place, in a format that machines readily understand, owned by the researcher. ORCID has created a model that balances self-assertion (researcher-entered data) with trusted assertion (data pushed by verified member organizations) to maintain a layered trust (Meadows et al., 2019).

### 4.2 Integration with Publishers, Funders, and Universities

ORCID gains greater value when integrated with publisher systems. During the manuscript submission process, major publishers such as Elsevier, Springer Nature, Wiley, and Taylor & Francis collect ORCID iDs, embedding them in the article's metadata and CrossRef record. A DOI-identified publication directly links to its ORCID-identified authors, a fundamental building block of the scholarly graph.

Funding agencies were among the most powerful drivers of ORCID adoption. The UKRI (UK Research and Innovation), NIH (National Institutes of Health), ERC (European Research Council), and Wellcome Trust have either a strong encouragement or requirement for ORCID in grant applications, enabling automated tracking of funded research outcomes. A study of 26 research institutions in Germany in 2022 revealed a broad range of ORCID adoption from 41% to 89%, which varied according to institutional policies and integration into Human Resources systems. More and more universities are adopting ORCID into their systems. Using the member API in ORCID, institutional ORCID programs facilitate automated synchronization of employment records; linking deposits in institutional repositories to researcher profiles; and population of CRIS systems, among other things. British Columbia and Strathclyde universities are quoted as exemplars of an ORCID that is integrated institution-wide. The implementation reduced burden on administrators and improved reporting accuracy (Meadows et al., 2019).

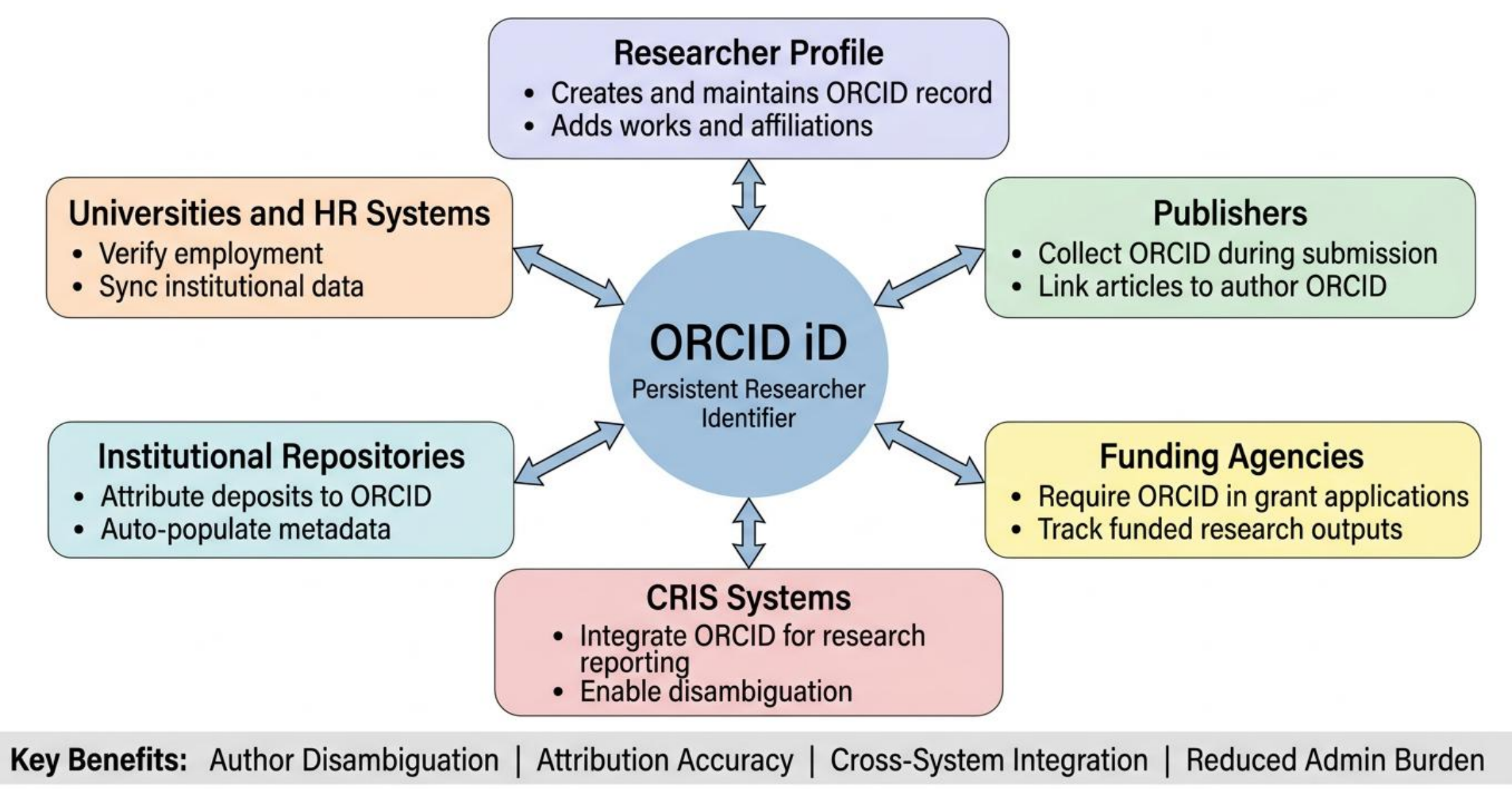

**Figure 3:** ORCID Researcher Identity Management Workflow

**Source(s):** Created by the author

## 5. ROR for Institutional Identity and Research Governance

### 5.1 Standardizing Institutional Affiliation

The metadata of institutional affiliations vary quite a lot. A single university may show up in publication records with dozens of name versions, abbreviated, translated, historic, and informal, making it difficult to accurately attribute research output to institutions. ROR's solution to this is the provision of a single open persistent identifier for each research organization, cross-linked to related identifiers including ISNI, Wikidata and the now-defunct GRID (Lammey, 2020).

Affiliation strings are automatically normalized by publishers and preprint servers through ROR adoption. Once a ROR ID is captured in a manuscript submission system, downstream systems (like CrossRef metadata, institutional repositories, CRIS systems, analytics, and so on) can all resolve to the same institutional record. This removes the burden of manually curating large volumes of bibliometric data, which has traditionally consumed significant library staff time and led to misallocation errors in research performance data.

### 5.2 Research Analytics, Rankings, and Funding Governance

Attributing an institution accurately is fundamental to research analytics and university rankings. The Times Higher Education World University Rankings and Leiden Ranking make use of bibliometric data that is only as accurate as the affiliation metadata that underpins it. An improvement in the reliability of institutional output counts, collaboration maps and citation impact measures may result from ROR-enabled affiliation standardization (Lammey, 2020).

ROR offers a sustainable organizational identifier for research funding governance. This would link grant records with the outputs of the linked institutions. Funding gotcha's can find out which ROR-identified institutions produced the publications, datasets and patents resulting from specific grants. The European Commission’s OpenAIRE infrastructure that aggregates research outputs from across European funding programs has adopted ROR as its standard organizational identifier, permitting unprecedented-scale cross-border funding impact analysis (Manghi et al., 2020).

### 5.3 CRIS Systems and Institutional Repositories

Currently, CRIS is the main administrative infrastructure through which the universities manage and report their research activities. The ROR integration with CRIS platforms (Pure (Elsevier), Converis (Clarivate), and VIVO) allows systematic filling of organizational metadata, eliminating manual data entry and ensuring a more consistent institutional research profile. Likewise, institutional repositories using DSpace, EPrints, or Invenio can embed ROR identifiers into their deposit workflows, ensuring unambiguous attribution of every item to the producing institution.

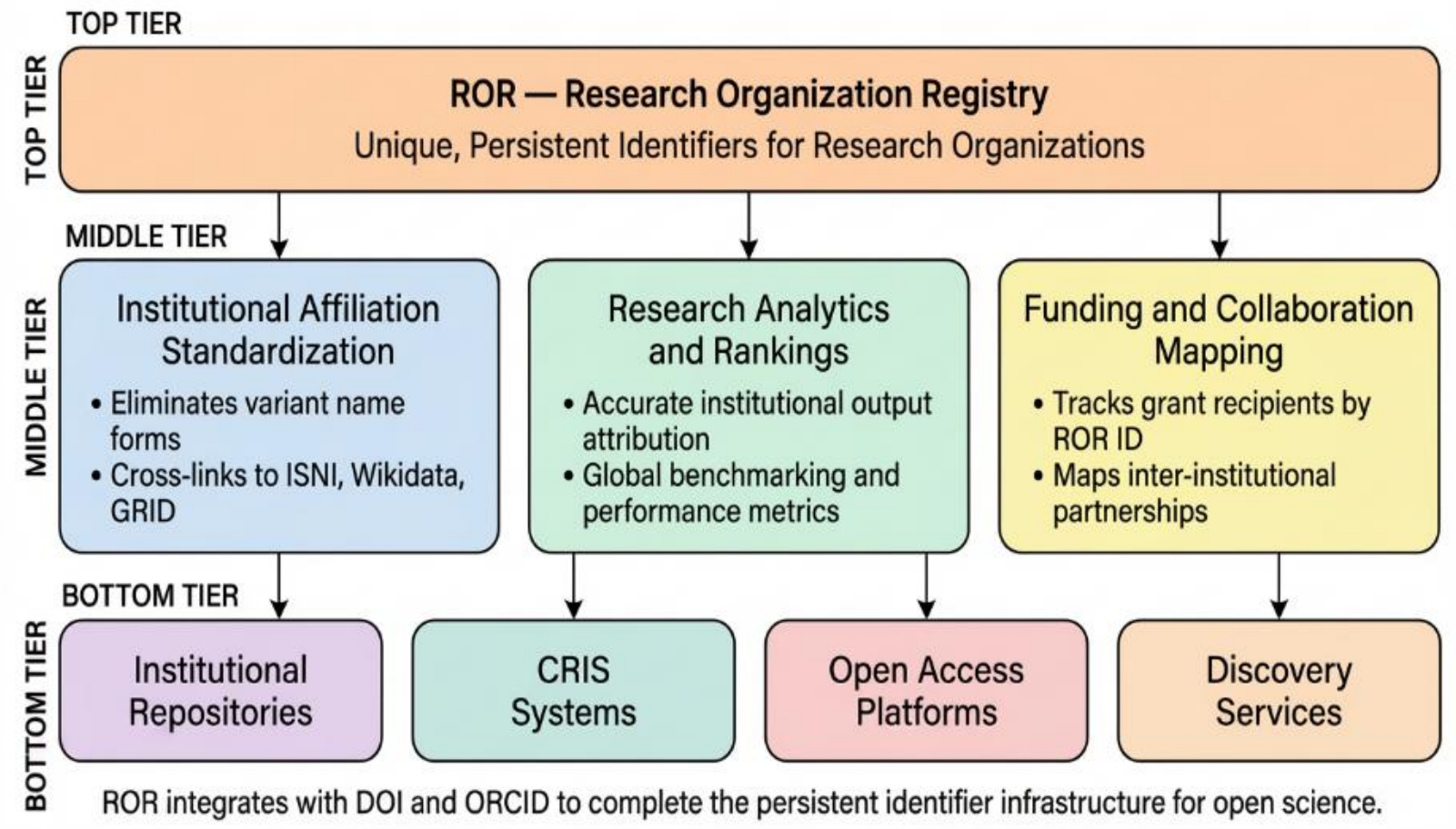

**Figure 4:** ROR: Standardizing Institutional Identity

**Source(s):** Created by the author

## 6. Interoperability in Libraries

### 6.1 PID Integration Across Library Systems

The operational impact of PIDs may only be realized when it is integrated across the entire spectrum of library and research information systems. Services like Ex Libris Primo, EBSCO Discovery Service and OCLC WorldCat are producing an increasing number of DOI, ORCID and ROR metadata. This improves search accuracy, enables faceted filtering by institution or researcher, and yields additional contextual information about retrieved items (Bilder et al., 2015). With rich PID metadata, a library catalogue is no longer merely a finding tool. It is now a repository in a global knowledge graph able to answer questions previously requiring sifting through systems.

DOIs are the main linking identifier used in many open access platforms, such as Europe PMC and CORE and registered repositories on OpenDOAR. Moreover, they are more and more integrating ORCID and ROR into their infrastructure to facilitate browsing of open-access content at the researcher and institution level. The "open by default" vision in Plan S and similar open access mandates will be supported by this integration, with PIDs providing the machine-readable infrastructure to verify compliance at scale. Institutions with metadata workflows that use PIDs are in a much better position to facilitate institutional open access reporting, lessen the administrative burden of compliance monitoring, and highlight the reach and impact of open access programs to institutional leadership.

### 6.2 APIs, Linked Data, and Machine-Readable Metadata

The ability of PIDs to work with each other relies on open and well-documented APIs. Library systems can programmatically query, retrieve, and update PID metadata using the RESTful APIs provided by CrossRef, ORCID, and ROR. Libraries can use CrossRef's metadata API to retrieve a full bibliographic record for any DOI. ORCID's member API allows for two-way data exchange with institutional systems. ROR's API provides an affiliation string match and access to Organizational records (Fenner et al. 2019). Other than APIs, PIDs are already naturally compatible with the principles of linked data and the semantic web. Since DOIs, ORCID IDs and ROR IDs have been expressed as HTTP URIs, they can be used as subjects and/or objects in RDF triples. Libraries can therefore be integrated into the wider Knowledge Graph, linking their holdings and researcher profiles to external resources such as Wikidata,

VIAF and disciplinary ontologies. The Scholix framework for data-literature linking and OpenCitations Corpus use DOIs as main linking mechanisms, showing how PID-based linked data can create a richly interlinked scholarly knowledge graph Heibi et al. (2019).

**7. Benefits for Institutions and Researchers**

The smart utilization of DOI, ORCID and ROR provides measurable benefits in various aspects of library and research management. Outputs with persistent, resolvable identifiers substantially improve the visibility and discoverability of research. Easy-to-use DOI-enabled publications are more reliably indexed, more accurately cited, and more widely distributed through aggregators. ORCID-linked researcher profiles ensure outputs are attributed correctly, no matter the name, while ROR-normalized affiliations ensure institutional outputs are fully captured in analytics systems.

Attribution clarification enhances research integrity. PIDs help avoid misattributions, double counting, and citation errors, which can distort research assessment. For institutions facing pressure to show research performance for funding and ranking, PID-enabled accuracy is not just useful but strategically essential. A great benefit that is often overlooked is administrative efficiency. Processing the data is done through automated feed between ORCID, ROR, institutional HR systems, CRIS systems, and repositories. Research at European universities shows that after the implementation of ORCID, the administrative burden facing researchers has fallen by 30-50%. Combining ORCID to identify co-authors and ROR to identify their institutions enables collaboration mapping and funding tracking. Research offices and funders can visualize global collaboration networks, identify productive partnerships, and target investment more. The Research Graph of OpenAIRE that combines DOI, ORCID, and ROR metadata from across the European research is an example of this capability at a continental scale (Manghi et al., 2020).

When libraries are able to query PID-enriched metadata, they can understand trends in institutional publishing. This power vastly improves data-driven decision making for library collection development, research support services, and open access strategy. Furthermore, libraries can also identify high-impact research areas. In addition, libraries can keep a check on compliance with open access mandates.

A key long-term benefit of investing in PID infrastructures resides in the way they will transform library services from reactive to proactive i.e. exploiting the insights from PIDs to anticipate researcher services rather than responding just to requests from researchers. Institutions that view PIDs as strategic data assets which carry institutional value rather than as housekeeping tasks are likely to be able to show value to institutional leadership and obtain resources for further development.

**8. Challenges and Barriers**

The presence of a transformative PID, which isn't happening due to barriers, is at the core of our mobilization theory. The two constraints that are most consistently recorded are low awareness and low PID literacy. Most researchers know about DOIs, but understanding of ORCID and ROR is much lower, especially amongst early career researchers and those in disciplines with weaker open access traditions (Wilsdon et al., 2021).

The assistive technology integration problems are greatly encountered by institutions that have old library management systems. Many libraries, particularly those in developing countries, lack the software development expertise needed to implement ORCID's member API or ROR's affiliations matching service. The use of library platforms worldwide has been hindered due to a lack of integration solutions.

The value of the PIDs is undermined due to the metadata quality issues. Research on CrossRef metadata has established high rates of missing and/or incorrect ORCID IDs and ROR affiliations recorded for published articles, due to inconsistent information collection by publishers and authors (Wilsdon et al., 2017). The reliability of PID-based analytics depends fundamentally on the quality of the underlying metadata.

A lack of policy and governance at the institutional level makes the adoption of PID dependent on the enthusiasm of the researcher. It remains voluntary and uneven adoption due to the absence of institutional mandates ORCID. Many nations, like Australia, are making concerted efforts to develop specific strategies around PID and have begun implementing them across a range of institutional entities.

Concerns about global equity are steeped. According to (Alperin et al., 2025), a mounting set of barriers limit the ability of Global South researchers and institutions to adopt PIDs. These barriers include lack of awareness and support, inadequate technical infrastructure, language barriers in PID documentation and support, and overlooked nature through which PID-based analytics will further concentrate research visibility in well-resourced Global North institutions. We need to build capacity, multi-lingual services and inclusive governance for the PID infrastructure.

**Table 2:** Challenges and Solutions in PID Implementation

| Challenge | Key Barriers | Impacts | Solutions | Recent Evidence (2023–2025) |
|---|---|---|---|---|
| Awareness & Literacy | Low PID knowledge among researchers; disciplinary variation | Incomplete coverage; resistance to adoption | Training programs; integration into research methods curricula; institutional champions | Czech 58.8% ORCID coverage (Dvořák et al., 2024); #PIDsLATAM23 training events (Bongiovani et al., 2023) |
| Technical Integration | Legacy systems; lack of turnkey solutions; limited IT staff | High costs; inconsistent implementations | Open-source tools; standardized APIs; shared infrastructure initiatives | UK PID-optimized workflow prototype (McNicholl et al., 2024); FAIRCORE4EOSC PIDGraph APIs (Vergoulis et al., 2025) |
| Metadata Quality | Missing/incorrect PIDs in publications; inconsistent formatting | Failed automated workflows; inaccurate analytics | Validation tools; publisher mandates; automated quality checking | 54.6% DOI coverage in OA diamond journals (Jamali et al., 2026); Nelson Memo re-curation mandates (Habermann, 2023) |
| Policy & Governance | No institutional PID policies; unclear mandates | Uneven adoption; limited institutional support | Institutional policies; funder mandates; national coordination | National Transportation Library PID mandate (2024); UK national PID strategy (McNicholl et al., 2024) |
| Resource Constraints | Limited budgets; competing priorities | Delayed implementations; manual processes | Phased approaches; shared services; external funding | FAIRCORE4EOSC shared infrastructure (Vergoulis et al., 2025); consortium models |
| Global Equity | Language barriers; infrastructure gaps in Global South | Exclusion of developing country researchers; biased analytics | Multilingual resources; capacity building; subsidized access | Latin America landscape analysis (Garduño-Magaña, 2024); AffilGood multilingual tools (Duran-Silva et al., 2024) |
| Privacy & Control | Concerns about data ownership; commercial use of PID data | Resistance; trust issues | Clear data governance policies; researcher control mechanisms | ORCID researcher-controlled profiles; 73% active maintenance in Spain (Arroyo-Machado et al., n.d.) |
| Adoption Incentives | Unclear benefits; administrative burden | Low voluntary adoption | Demonstrate tangible benefits; automate data flows; integrate into | 30–50% administrative burden reduction (Meadows et al., 2019); UBC ROR |

| | | | existing workflows | implementation (Barsky & Sprout, 2025) |
|---|---|---|---|---|

**9. Global Best Practices and Case Studies**

**9.1 European Leadership in PID Infrastructure**

Europe has been working intensively to coordinate PID infrastructure in this region. Since 2020, the German PID Network has combined the efforts of research institutions, libraries, and funders to achieve a greater alignment of PIDs in the German research area. In an audit of 26 member institutions in 2022, the use of ORCID varied widely across institutions; adoption was between 41% and 89%. Notably, the highest rates were found at institutions that had embedded ORCID into HR onboarding and CRIS workflows (Dvořák et al., 2024).

The Dutch universities' SURF cooperative has developed a common PID infrastructure. This is to integrate ORCID and ROR into the national CRIS ecosystem. Thus, to provide centralized technical support, which institutions cannot afford. This consortium model has gained almost complete coverage of ORCID at Dutch research universities and often gets quoted as a template for national PID coordination (Bilder et al., 2015).

The Research Excellence Framework (REF) in the UK has been a strong influence on PID use as institutions must report research outputs with correct attribution and affiliation information. The prospect of REF assessment has spurred universities in the UK to invest in ROR-normalized affiliation data and ORCID integration. The University of Strathclyde and University College London have documented institution-wide ORCID programmes (Meadows et al., 2019).

**9.2 Publisher and Platform Adoption**

Over the last five years, a major initiative of CrossRef, called the Metadata Plus Program, has enabled significant improvements in DOI metadata quality amongst CrossRef's more than 20,000 member publishers. CrossRef certification indicates publishers' commitment to achieve compliance. Metadata such as ORCID iDs and ROR affiliation, etc, should be complete and accurate. A CrossRef-certified metadata quality tier can help libraries vet the reliability of data. In 2023, Elsevier, Springer Nature, and Wiley introduced embedded ORCID registration features in their submission systems. Wiley states that over 70% of new submissions are registered by authors. The integration of DataCite with institutional repositories has widened DOI registration to beyond publishing. DataCite assigns DOIs for research data, software and grey literature in over 800 repositories worldwide, which means persistent identification is available for all research outputs. The European Open Science Cloud (EOSC) has made the use of DOI, ORCID, and ROR mandatory identifiers for all outputs deposited in EOSC-compliant repos (Manghi et al., 2020).

**9.3 Emerging Adoption in Latin America and Asia**

The adoption of regional PIDs in Latin America has relied most heavily on the SciELO network. The Scientific Electronic Library Online (SciELO) which offers open access to Brazilian, Argentine, Chilean, etc., journals, is more frequently registering DOI and collecting ORCID on its platform, allowing Global South journals to engage with the global PID ecosystem. Researchers applying for support from Brazil's FAPESP funding agency will now have to provide ORCID as part of their applications. This policy mirrors mandates from European funders and aims to increase registration rates among researchers (Alperin et al., 2025).

In Asia, national universities in Japan and Peking University, Taiwan, have implemented institutional ORCID programs intended to work with their research information system, showing PID implementation is not the prerogative of Western research. China's National Natural Science Foundation is investigating the possibility of integrating ORCID in its grant management process, which might speed up the ORCID coverage globally.

## 10. Future Trends

### 10.1 AI-Driven Research System and Smart Libraries

The utilization of PIDS in library and information systems and in research systems is undergoing a change due to AI. Machine learning models that utilize PID-enriched metadata are superior to rule-based systems in disambiguating affiliations, predicting citations, and identifying emerging fronts. So, the emerging tools like OpenAlex, which is now a fully open scholarly knowledge graph built on DOI, ORCID, and ROR data. AI applications can leverage the PID infrastructure to provide research intelligence at scale without licensing a commercial database (Priem et al., 2022).

In the future, smart library systems will leverage PID-based knowledge graphs to provide personalized discovery experiences, auto-update their institute's research profiles, and monitor open access compliance at the point of engagement. Through the incorporation of PIDs within natural language processing, libraries will shift from keyword search to concept search, leading users to researchers, institutions, datasets and funding streams and not just documents relevant to their topic.

### 10.2 Expanding the PID Ecosystem

The PID ecosystem is extending beyond DOI, ORCID and ROR. RAiD (Research Activity Identifier) generates persistent identifiers of research projects allowing the linking of grants, outputs, researchers and institutions throughout the lifecycle of the project. The International Geo Sample Number cites the Persistent Identifier (PID) infrastructure in fieldwork, based disciplines.  Researchers are developing instrument PIDs are under development to identify the specific equipment being used in research, to ensure reproducibility and provide equipment-level analytics (Fenner et al., 2019).

The PIDs in question will form an increasingly richer scholarly knowledge graph when they connect with DOI, ORCID, and other ROR. Those libraries that have built PID-aware systems and PID-aware workflows will be well-positioned to accommodate new identifier types as and when they reach maturity; others that have not invested in PID infrastructure will face compounding integration problems.

### 10.3 Open Science and PID-Based Policy Compliance

Over the last ten years, a coalition of countries and international organizations have been supporting the global open science movement. The European open access mandate Plan S requires funder DOIs for all publications it funds and is moving to requiring ORCID and ROR in funder metadata. According to the UNESCO Recommendation on Open Science (UNESCO, 2021), persistent identifiers are an essential component of open science. Due to the global application of these policies, the use of PID will grow to be a fund-eligibility criteria instead of best practices.

## 11. Recommendations

It is recommended that libraries assign DOIs to local research outputs; integrate ORCID with repositories and research systems; and adopt ROR to accurately associate research outputs with their institution (Barsky & Sprout, 2025; Dvořák et al., 2024; Habermann, 2023). Training for staff, researchers and students needs to be conducted to enhance everyday use of these identifiers and strengthen discovery of research outputs. New Universities and colleges should ask for ORCID upon onboarding, use identifiers during research workflow, and spend on systems able DOI, ORCID, and ROR. Involvement in national identifier networks and policies at the institution level would enhance reporting, reduce administrative burden and increase visibility of research (Arroyo-Machado et al., n.d.; McNicholl et al., 2024). Funding agencies and policymakers should require research they fund to have identifiers, support national infrastructures, and incorporate them into open science policies. Garduño-Magaña, 2024; Vergoulis et al. 2025 argued that developing regions need financial and training support to participate globally. Library experts must gain metadata standards higher-level skills, APIs and research analytics capabilities, advocate for identifiers, and actively participate in Crossref, ORCID, DataCite and global communities. The involvement of libraries in present-day scholarly communication will certainly expand (French & Buttrick, 2023; Priem et al., 2022).

## 12. Conclusion

DOI, ORCID and ROR are important building blocks for an efficient open library ecosystem. The provision of stable machine-readable identification of research objects, researchers and organizations transforms fragmented metadata into a coherent scholarly knowledge graph. It helps with exploration and evaluation while enabling opening science globally. The evidence discussed in this review shows the promising ability of persistent identifiers (PIDs) and that substantial progress must be made. Their use remains uneven, metadata quality is inconsistent, and equity gaps between the Global North and South persist. To fulfill the full promise of PID infrastructure, concerted activity is essential by libraries, universities, funders, publishers, and policymakers. Institutions such as libraries are part of this infrastructure as beneficiaries and caretakers and their involvement in adopting, supporting, and researching persistent identifiers (PIDs) is vital for building the open interoperable scholarly ecosystem contemporary research requires.